\documentclass[prd,twocolumn,showpacs]{revtex4}
\usepackage{graphicx}
\usepackage{latexsym}
\usepackage{amsmath}
\begin{document}
\title{Reconstruction of the deceleration parameter and the equation of state of dark energy}
\author{Yungui Gong}
\email{yungui_gong@baylor.edu}
\affiliation{College of Electronic Engineering, Chongqing
University of Posts and Telecommunications, Chongqing 400065,
China}
\affiliation{CASPER, Physics Department, Baylor University,
Waco, TX 76798, USA}
\author{Anzhong Wang}
\email{anzhong_wang@baylor.edu}
\affiliation{CASPER,
Physics Department, Baylor University, Waco, TX 76798, USA}
\begin{abstract}
The new 182 gold supernova Ia data, the baryon acoustic oscillation measurement and
the shift parameter determined from the Sloan Digital Sky Survey and the three-year
Wilkinson Microwave Anisotropy Probe data are combined to reconstruct the
dark energy equation of state parameter $w(z)$ and the deceleration parameter $q(z)$. We find that
the strongest evidence of acceleration happens around the redshift $z\sim 0.2$ and the stringent
constraints on $w(z)$ lie in the redshift range $z\sim 0.2-0.5$. At the sweet spot,
$-1.2<w(z)<-0.6$ for the dark energy parametrization $w(z)=w_0+w_a z/(1+z)^2$ at
the $3\sigma$ confidence level. The transition redshift $z_t$
when the Universe underwent the transition from deceleration to acceleration
is derived to be $z_t=0.36^{+0.23}_{-0.08}$. The combined data is also applied
to find out the geometry of the Universe, and we find that at
the $3\sigma$ confidence level, $|\Omega_k|\alt 0.05$  for the simple one parameter dark energy model,
and $-0.064<\Omega_k<0.028$ for the $\Lambda$CDM model.

\end{abstract}
\preprint{astro-ph/0612196}
\maketitle

\section{Introduction}

The discovery of the accelerated expansion of the Universe by the
supernova (SN) Ia observations \cite{agr98} imposes a big challenge and provides
opportunities to theoretical physics. The more
accurate SN Ia data \cite{riess,astier,riess06}, the three-year
Wilkinson Microwave Anisotropy Probe (WMAP3) data \cite{wmap3}, and the Sloan
Digital Sky Survey (SDSS) data \cite{sdss} tell us that the Universe
is almost spatially flat, and dark energy (DE) with negative pressure
contributes about 72\% of the matter content of the Universe. Although the
existence of DE were verified by different observations, the
nature of DE is still a mystery to us. For a review of DE models,
one may refer to Ref. \cite{DE}.

Many parametric and non-parametric model-independent methods were proposed
to study the evolutions of the deceleration parameter $q(z)$, the DE density
$\Omega_{DE}(z)$, the DE equation of state (EoS) $w(z)$, and the geometry
of the Universe \cite{virey,sturner,gong06,astier01,huterer,weller,alam,gong04,gong05,lind,jbp,par1,par2,
par3,par4,par5,gong04a,wang05,jbp05, nesseris6a,berger,sahni06,lihong,yun06,evldh,saini,jbp06,nesseris05,nesseris6b}.
In the reconstruction of $q(z)$, it was found that the strongest evidence of acceleration happens at
redshift $z\sim 0.2$ \cite{virey,sturner,gong06}, and the evidence of the current acceleration is not
very strong \cite{sturner} and model dependent \cite{gong06}. Previous studies
on the reconstruction of $w(z)$ also showed that the stringent constraint on $w(z)$, or the sweet spot,
happened around redshift $z\sim 0.2-0.5$ \cite{astier01,huterer,weller,alam,gong04,gong05}. As
Riess {\it et al}. pointed out, the use of additional parameters to reconstruct
$w(z)$ does not provide a statistically significant improvement on the fit of the
redshift-magnitude relation, so we discuss one- and two-parameter models only.
In this work, we first use the simple two-parameter model $
q(z)=1/2+(q_1 z+q_2)/(1+z)^2$ \cite{gong06}
to reconstruct $q(z)$, then we use the three popular two-parameter
models $w(z)=w_0+w_a z/(1+z)$ \cite{lind}, $w(z)=w_0+w_a z/(1+z)^2$ \cite{jbp}
and $\Omega_{DE}=1-\Omega_m-A_1-A_2+A_1(1+z)+A_2(1+z)^2$ \cite{alam} to reconstruct $w(z)$.
The purpose of this work is to see if the stringent constraints on $q(z)$ and $w(z)$
still happen around $z\sim 0.2-0.5$ when we use the new 182 gold SN Ia data compiled in \cite{riess06}.
The geometry of the Universe is also discussed by fitting the simple one-parameter model
$w(z)=w_0\exp[z/(1+z)]/(1+z)$ \cite{gong05} to the combined SN Ia, SDSS and WMAP3 data.

This paper is organized as follows. In section II,
we study the property of $q(z)$ by fitting the parametrization $q(z)=1/2+(q_1 z+q_2)/(1+z)^2$
to the new 182 gold SN Ia data compiled in \cite{riess06}.
In section III, we apply the popular parameterizations $w(z)=w_0+w_a z/(1+z)$,
$w(z)=w_0+w_a z/(1+z)^2$ and $\Omega_{DE}=1-\Omega_m-A_1-A_2+A_1(1+z)+A_2(1+z)^2$ to study the
evolutions of DE EoS. The baryon acoustic oscillation (BAO) measurement from SDSS and
the shift parameter determined from WMAP3 data combined with the new 182 gold SN Ia data
are used in our analysis. In section IV, we fit the simple one-parameter representation
$w(z)=w_0\exp[z/(1+z)]/(1+z)$ to the combined SN Ia, SDSS and WMAP3 data to
obtain the geometry of the Universe. Note that
the simple one-parameter model fits the observational data as well as the two-parameter
models do. In section V, we conclude the paper with some discussions.

\section{Reconstruction of the deceleration parameter}

The Hubble parameter $H(t)=\dot{a}/a$ and the deceleration
parameter $q(t)=-\ddot{a}/(aH^2)$ are related by the following equation,
\begin{equation}
\label{hubq}
H(z)=H_0\exp \left[\int^z_0 [1+q(u)]d\ln(1+u)\right],
\end{equation}
where the subscript 0 means the current value of the variable.
If a function of $q(z)$ is given, then we can find the evolution of the Hubble parameter.
For the flat $\Lambda$CDM model, $q(z)=[\Omega_m(1+z)^3-2(1-\Omega_m)]/2[\Omega_m(1+z)^3+1-\Omega_m]$.
In this section, we use the simple two-parameter function \cite{gong06}
\begin{equation}
\label{qmod2}
q(z)=\frac{1}{2}+\frac{q_1 z+q_2}{(1+z)^2},
\end{equation}
to reconstruct the evolution of $q(z)$. Note that $q_0=1/2+q_2$, and
$dq/dz|_{z=0}=q_1-2q_2$,
so the parameter $q_2$ gives the value of $q_0$.
At early times, $z\gg 1$, $q(z)\rightarrow 1/2$.
Substitute Eq. (\ref{qmod2}) into Eq. (\ref{hubq}),
we get
\begin{equation}
\label{hubsl2}
H(z)=H_0(1+z)^{3/2}\exp\left[\frac{q_2}{2}+\frac{q_1 z^2-q_2}{2(1+z)^2}\right].
\end{equation}
Since the expression for the Hubble parameter is explicit, so we can think that
we are actually modelling $H(z)$ instead of $q(z)$.

The parameters $q_1$ and $q_2$ in the model are determined by minimizing
\begin{equation}
\label{chi}
\chi^2=\sum_i\frac{[\mu_{obs}(z_i)-\mu(z_i)]^2}{\sigma^2_i},
\end{equation}
where the extinction-corrected distance modulus $\mu(z)=5\log_{10}[d_L(z)/{\rm Mpc}]+25$,
$\sigma_i$ is the total uncertainty in the SN Ia data,
and the luminosity distance is
\begin{equation}
\label{lum}
d_L(z)=(1+z)\int_0^z\frac{dz'}{H(z')}.
\end{equation}

\begin{figure}
\centering
\includegraphics[width=8cm]{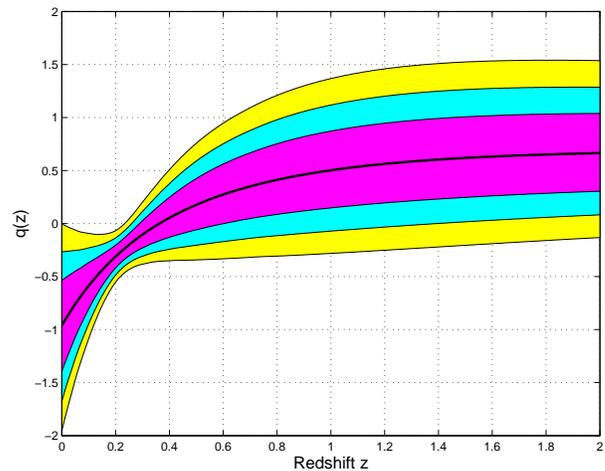}
\caption{The evolution of $q(z)=1/2+(q_1 z+q_2)/(1+z)^2$ by
fitting it to the 182 gold SN Ia data. The solid line is drawn by
using the best fit parameters. The shaded areas show the $1\sigma$, $2\sigma$
and $3\sigma$ errors.}
\label{fig1}
\end{figure}

Fitting the model to the 182 gold SN Ia data, we get $\chi^2=156.25$, $q_1=1.47^{+1.89}_{-1.82}$
and $q_2=-1.46\pm 0.43$, here the given error is the $1\sigma$ error.
For comparison, we also fit the $\Lambda$CDM model to the 182 gold SN Ia data and find
that $\chi^2=156.16$, $\Omega_m=0.48^{+0.13}_{-0.15}$ and $\Omega_k=-0.44^{+0.43}_{-0.36}$.
So the simple two-parameter model of $q(z)$ fits the SN Ia data as well as the $\Lambda$CDM model does.
By using the best fitting results, we plot the evolution of $q(z)$ in Fig. \ref{fig1}.
From Fig. \ref{fig1}, we see that $q(z)<0$ for $0\leq z\alt 0.2$ at the $3\sigma$ confidence level.
This result is consistent with previous analysis by using the 157 gold SN Ia data \cite{gong06}.
It is also interesting to note that the stringent constraint on $q(z)$ happens around the redshift
$z\sim 0.2$. One may think perhaps there are more SN Ia data around this redshift. On the contrary, less
SN Ia data is around $z=0.2$. In table \ref{sntab}, we list the number $N$ of SN Ia in a given
redshift range for the 182 gold SN Ia data. The behavior
was also found in \cite{gong06,astier01,huterer,weller,alam,gong04,gong05} in the fitting of the EoS of DE for
a variety of models. This may suggest that the behavior of DE can be better constrained
in the redshift range $0.1\alt z \alt 0.6$. The sweet spot can be estimated from the covariant
matrix of errors, which is the inverse of the Fisher matrix in the linear approximation \cite{astier01,huterer,weller}.
The Fisher matrix is estimated to be $F_{11}=2.37$, $F_{12}=F_{21}=9.55$, and $F_{22}=43.9$. By
choosing $\alpha_1=q_1$ and $\alpha_2=(F_{12}/F_{22}) q_1+q_2=0.2175 q_1+q_2$, the Fisher matrix
becomes diagonal, $\alpha_1$ and $\alpha_2$ are uncorrelated, and the errors of $\alpha_1$
and $\alpha_2$ are $\sigma^2(\alpha_1)=F_{22}/(F_{11}F_{22}-F_{12}^2)$ and
$\sigma^2(\alpha_2)=F^{-1}_{22}$ \cite{dje}. In terms of $\alpha_1$ and $\alpha_2$, we get
\begin{equation}
q(z)=\frac{1}{2}+\frac{\alpha_1(z-0.2175)+\alpha_2}{(1+z)^2}.
\end{equation}
Now the sweet spot can be estimated from the following equation
\begin{equation}
\frac{2[\sigma^2(\alpha_2)+\sigma^2(\alpha_1)(z-0.2175)^2]}{1+z}=\sigma^2(\alpha_1)(z-0.2175).
\end{equation}
The sweet spot is estimated to be $z\simeq 0.2175$ since $\sigma^2(\alpha_2)=F^{-1}_{22}\sim 0$.
For a general model $w(z)=w_1+w_2 f(z)$ with arbitrary function $f(z)$, the sweet spot is determined similarly from the equation
$f(z)=F_{12}/F_{11}$.

%\begin{widetext}
\begin{table*}
\begin{center}
\caption{The distribution of the gold SN Ia data}
\label{sntab}
\begin{tabular}{|c|c|c|c|c|c|c|c|c|c|c|c|}
  \hline
$z$ & $<0.1$ & 0.1-0.2 & 0.2-0.3 & 0.3-0.4 & 0.4-0.5 & 0.5-0.6 & 0.6-0.7 & 0.7-0.8 & 0.8-0.9 & 0.9-1.0& $>1.0$\\
  \hline
$N$& 36 & 4 & 5 & 12 & 31 & 22 & 16 & 11 & 17 & 12 & 16
\\\hline
\end{tabular}
\end{center}
\end{table*}
%\end{widetext}

\section{Dark Energy Parametrization}

In this section, we use the observational data to reconstruct the EoS of DE.
For simplicity, we consider the spatially flat case, $k=0$.
In addition to the SN Ia data, we also use the distance parameter
\begin{equation}
\label{paraa}
A=\frac{\sqrt{\Omega_{m}}}{0.35}\left[\frac{0.35}{E(0.35)}\left(\int^{0.35}_0\frac{dz}{E(z)}\right)^2\right]^{1/3},
\end{equation}
measured from the SDSS data to be $A=0.469(0.95/0.98)^{-0.35}\pm 0.017$ \cite{sdss,wmap3},
and the shift parameter  \cite{yun06}
\begin{equation}
\label{shift}
\mathcal{R}=\sqrt{\Omega_{m}}\int_0^{z_{ls}}\frac{dz}{E(z)}=1.70\pm 0.03,
\end{equation}
where $E(z)=H(z)/H_0$ and $z_{ls}=1089\pm 1$.

\begin{figure}[htp]
\centering
\includegraphics[width=8cm]{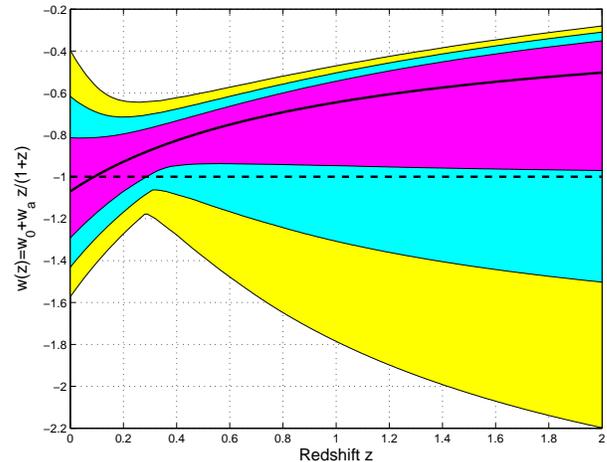}
\caption{The evolution of $w(z)$ by fitting the model $w(z)=w_0+w_a z/(1+z)$ to the
observational data.
The solid line is drawn by using the best fit parameters. The shaded areas show the $1\sigma$, $2\sigma$
and $3\sigma$ errors.}
 \label{fig2}
\end{figure}

\begin{figure}[htp]
\centering
\includegraphics[width=8cm]{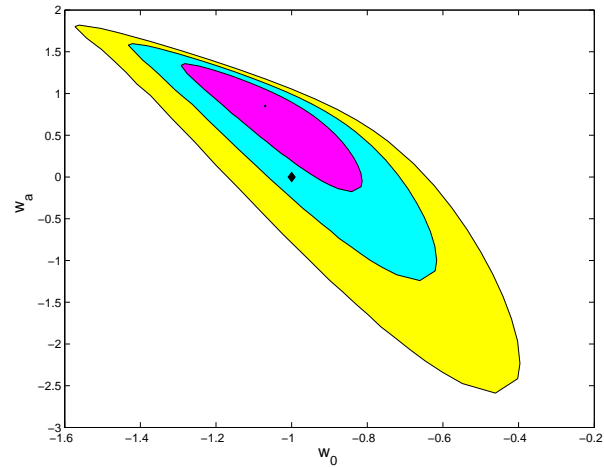}
\caption{The $1\sigma$, $2\sigma$ and $3\sigma$ contour plots of $w_0$ and $w_a$ for the
model $w(z)=w_0+w_a z/(1+z)$. The diamond denotes
the point corresponding to the cosmological constant.}
 \label{fig3}
\end{figure}

The first DE parametrization we consider is \cite{lind}
\begin{equation}
\label{lind}
w(z)=w_0+\frac{w_a z}{1+z}.
\end{equation}
The dimensionless DE density is
\begin{equation}
\label{deneq}
\Omega_{DE}(z)=\Omega_{DE0}(1+z)^{3(1+w_0+w_a)}\exp[-3w_a z/(1+z)].
\end{equation}
This parametrization can be thought as the parametrization of the DE density instead
of $w(z)$. By fitting this model to the observational data, we find that
$\chi^2=158.07$, $\Omega_m=0.29\pm 0.04$, $w_0=-1.07^{+0.33}_{-0.28}$ and $w_a=0.85^{+0.61}_{-1.38}$.
Compared with previous fitting results \cite{gong05},
the current data makes a little improvement on the constraint of $w_a$.
The evolution of $w(z)$ is plotted in Fig. \ref{fig2} and the contours of $w_0$
and $w_a$ are shown in Fig. \ref{fig3}. From Fig. \ref{fig2},
we see that at the $3\sigma$ confidence level, $w(z)<0$ for $z<2$, $w(z)$ crosses
the $-1$ barrier around $z\sim 0.1$, and the
stringent constraint on $w(z)$ happens around $z\sim 0.3$. From the Fisher matrix estimation,
we get $z/(1+z)=F_{12}/F_{11}=0.32$, so the sweet spot is around $z=0.47$. This estimation
is quite different from what we get, and the main reason is that the distribution of $w_a$ is highly
non-Gaussian. From Fig. \ref{fig3}, we see that the cosmological constant is more than $1\sigma$
away from the best fit result.

\begin{figure}[htp]
\centering
\includegraphics[width=8cm]{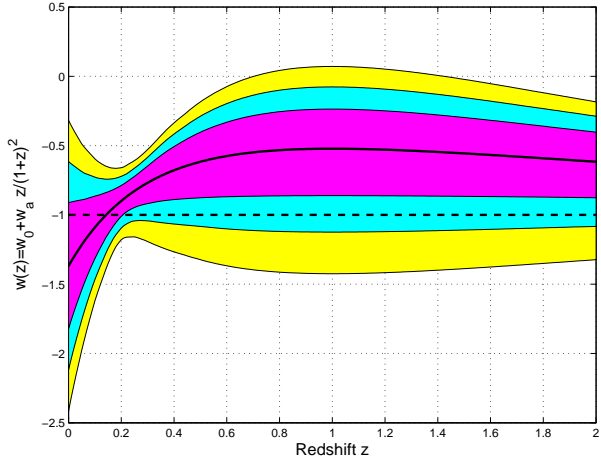}
\caption{The evolution of $w(z)$ by fitting the model $w(z)=w_0+w_a z/(1+z)^2$ to the
observational data.
The solid line is drawn by using the best fit parameters. The shaded areas show the $1\sigma$, $2\sigma$
and $3\sigma$ errors.}
\label{fig4}
\end{figure}

\begin{figure}[htp]
\centering
\includegraphics[width=8cm]{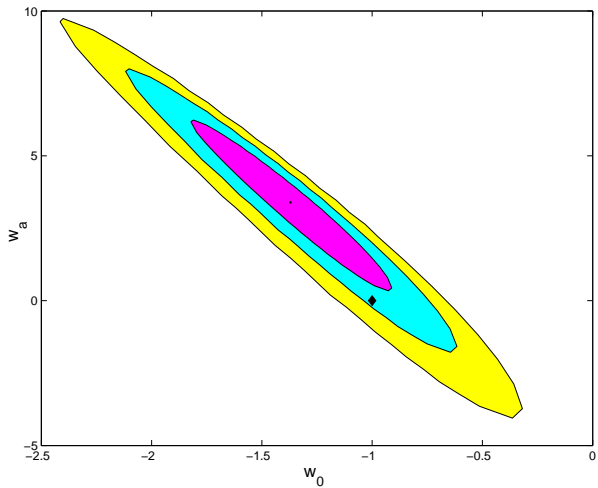}
\caption{The $1\sigma$, $2\sigma$ and $3\sigma$ contour plots of $w_0$ and $w_a$
for the model $w(z)=w_0+w_a z/(1+z)^2$. The diamond denotes
the point corresponding to the cosmological constant.}
 \label{fig5}
\end{figure}

The second DE parametrization we consider is \cite{jbp}
\begin{equation}
\label{wzeq}
w(z)=w_0+\frac{w_a z}{(1+z)^2}\ .
\end{equation}
The dimensionless DE density is
\begin{equation}
\label{deneq1}
\Omega_{DE}(z)=\Omega_{DE0}(1+z)^{3(1+w_0)}\exp\left[3w_a z^2/2(1+z)^2\right].
\end{equation}
Again this parametrization can also be thought as the parametrization of the DE density.
By fitting this model to the observational data, we find that
$\chi^2=157.11$, $\Omega_m=0.28^{+0.04}_{-0.03}$, $w_0=-1.37^{+0.58}_{-0.57}$ and $w_a=3.39^{+3.51}_{-3.93}$.
These constraints are almost at the same level as previous results in \cite{gong05}.
The evolution of $w(z)$ is plotted in Fig. \ref{fig4} and the contours of $w_0$
and $w_a$ are shown in Fig. \ref{fig5}. From Fig. \ref{fig4},
we see that at the $3\sigma$ confidence level, $w(z)<0$ for $z<0.7$, w(z) crosses the
$-1$ barrier around $z\sim 0.15$, and the
stringent constraint on $w(z)$ happens around $z\sim 0.2$. The sweet spot is estimated to be
$z=0.2353$ from the equation $z/(1+z)^2=F_{12}/F_{11}=0.1542$. From Fig. \ref{fig5},
we see that the $\Lambda$CDM model is almost $2\sigma$ away from the best fit result.

\begin{figure}[htp]
\centering
\includegraphics[width=8cm]{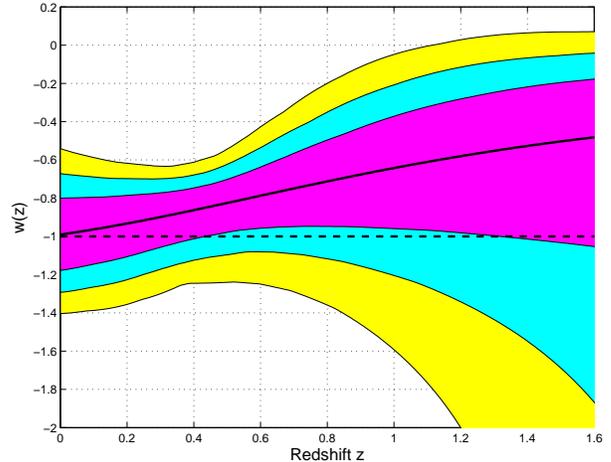}
\caption{The evolution of $w(z)$ by fitting the model
$\Omega_{DE}(z)=1-\Omega_m-A_1-A_2+A_1(1+z)+A_2(1+z)^2$ to the observational data.
The solid line is drawn by using the best fit parameters.
The shaded areas show the $1\sigma$, $2\sigma$
and $3\sigma$ errors.}
\label{fig6}
\end{figure}

\begin{figure}[htp]
\centering
\includegraphics[width=8cm]{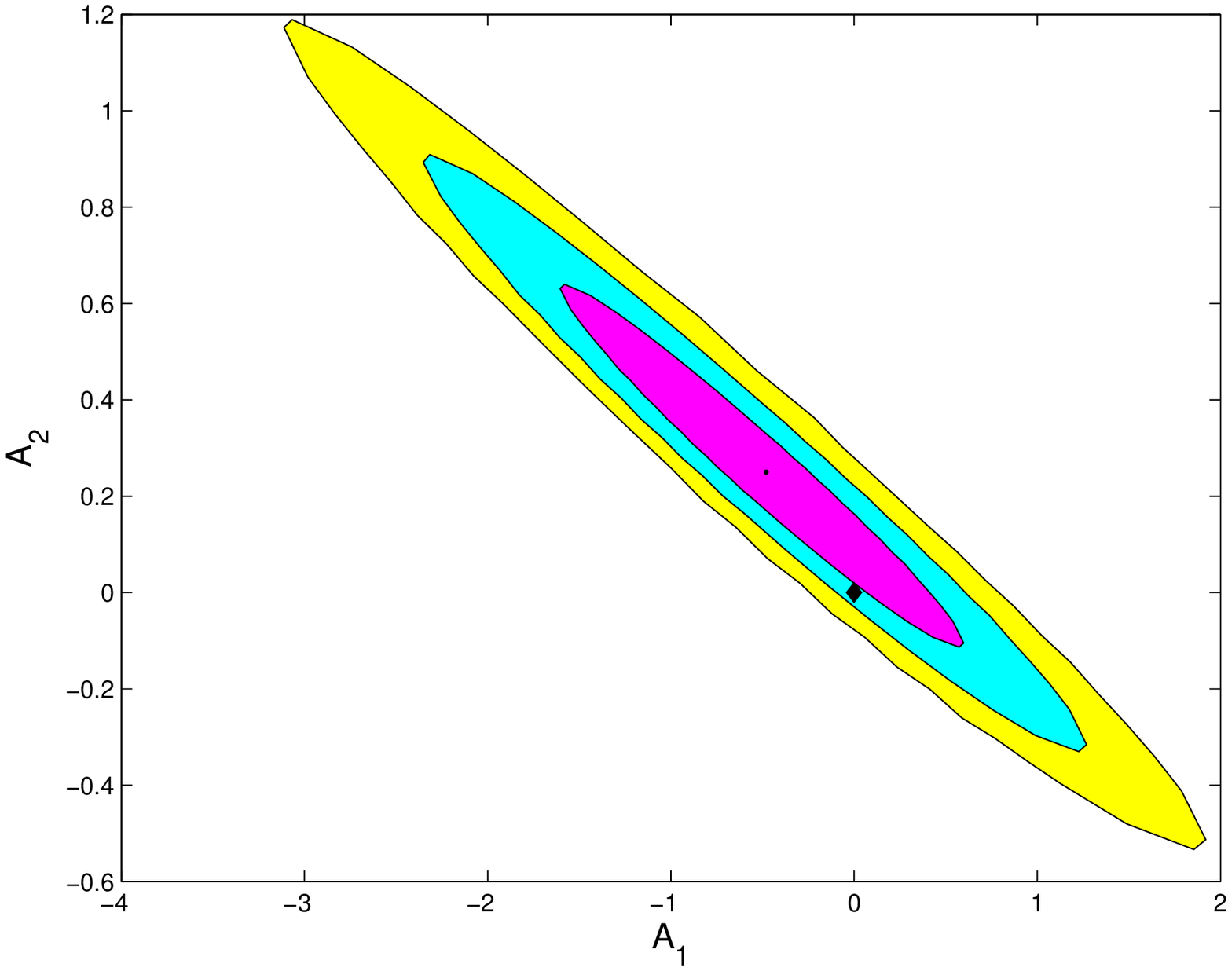}
\caption{The $1\sigma$, $2\sigma$ and $3\sigma$ contour plots of $A_1$ and $A_2$. The diamond denotes
the point corresponding to the cosmological constant.}
\label{fig7}
\end{figure}

The last model we consider is \cite{alam}
\begin{equation}
\label{poly2}
\Omega_{DE}(z)=A_1(1+z)+A_2(1+z)^2+1-\Omega_m-A_1-A_2.
\end{equation}
The EoS parameter $w(z)$ is
\begin{equation}
\label{wzpoly2}
w(z)=\frac{1+z}{3}\frac{A_1+2A_2(1+z)}{\Omega_{DE}(z)}-1.
\end{equation}
The cosmological constant corresponds to $A_1=A_2=0$.
By fitting this model to the observational data, we find that
$\chi^2=158.48$, $\Omega_m=0.30\pm 0.04$, $A_1=-0.48^{+1.36}_{-1.47}$ and $A_2=0.25^{+0.52}_{-0.45}$.
The evolution of $w(z)$ is plotted in Fig. \ref{fig6} and the contours of
$A_1$ and $A_2$ are shown in Fig. \ref{fig7}. From Fig. \ref{fig6},
we see that at the $3\sigma$ confidence level, $w(z)<0$ for $z<1.1$ and the
stringent constraint on $w(z)$ happens around $z\sim 0.4$.
From Fig. \ref{fig7}, we see that the $\Lambda$CDM model is more than
$1\sigma$ away from the best fit result.
Comparing the value of $\chi^2$ of the three models we considered,
we find that the second model fits a little bit better
than the other two models do.

\section{The geometry of the Universe}

In this section, we use the observational data to find out the geometry of our universe. When $k\neq 0$,
the luminosity distance becomes
\begin{eqnarray}
\label{lumdis}
d_{\rm L}(z)=\frac{1+z}{H_0\sqrt{|\Omega_{k}|}} {\rm
sinn}\left[\sqrt{|\Omega_{k}|}\int_0^z
\frac{dz'}{E(z')}\right],
\end{eqnarray}
where ${\rm sinn}(\sqrt{|k|}x)/\sqrt{|k|}=\sin(x)$, $x$, $\sinh(x)$ if $k=1$, 0, $-1$,
the parameter
$A$ becomes
\begin{equation}
\label{para1}
A=\frac{\sqrt{\Omega_{m}}}{0.35}\left[\frac{0.35}{E(0.35)}\frac{1}{|\Omega_{k}|}{\rm
sinn}^2\left(\sqrt{|\Omega_{k}|}\int_0^{0.35}
\frac{dz}{E(z)}\right)\right]^{1/3},
\end{equation}
and the shift parameter becomes
\begin{equation}
\label{shift1}
\mathcal{R}=\frac{\sqrt{\Omega_{m}}}{\sqrt{|\Omega_{k}|}}{\rm
sinn}\left(\sqrt{|\Omega_{k}|}\int_0^{z_{ls}}\frac{dz}{E(z)}\right).
\end{equation}

\begin{figure}[htp]
\centering
\includegraphics[width=8cm]{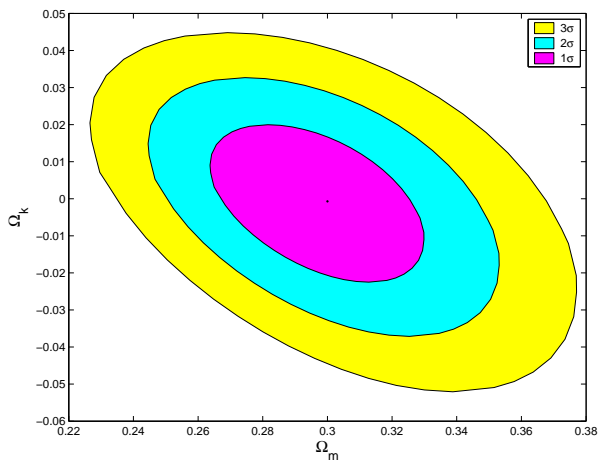}
\caption{The $1\sigma$, $2\sigma$ and $3\sigma$ contour plots of
$\Omega_{m}$ and $\Omega_{k}$ for the parametrization
$w(z)=w_0\exp[z/(1+z)]/(1+z)$.}
\label{fig8}
\end{figure}

\begin{figure}[htp]
\centering
\includegraphics[width=8cm]{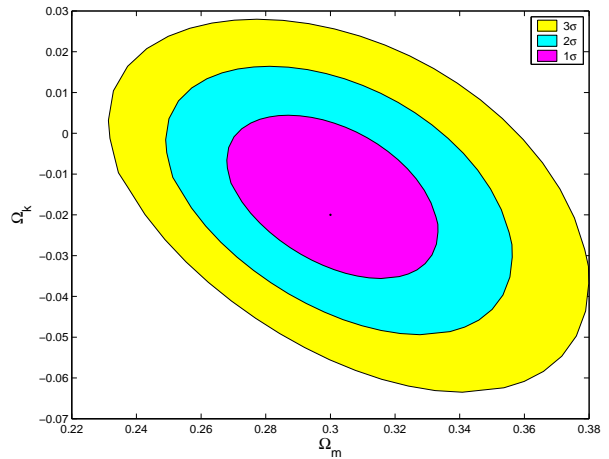}
\caption{The $1\sigma$, $2\sigma$ and $3\sigma$ contour plots of
$\Omega_{m}$ and $\Omega_{k}$ for the $\Lambda$CDM model.}
\label{fig9}
\end{figure}

To fit the observational data, we consider the one parameter DE parametrization \cite{gong05}
\begin{equation}
\label{1pnb}
w(z)=\frac{w_0}{1+z}e^{z/(1+z)}.
\end{equation}
During both the early and future epoches, $w(z)\rightarrow 0$. The DE density is
\begin{equation}
\label{dens1pb}
\Omega_{DE}=\Omega_{DE0}(1+z)^3\exp\left(3\omega_0
e^{z/(1+z)}-3\omega_0\right).
\end{equation}
By fitting this model to the observational data, we find that $\chi^2=158.85$,
$\Omega_m=0.30\pm 0.04$, $\Omega_k=-0.0007^{+0.032}_{-0.03}$ and $w_0=-0.93^{+0.17}_{-0.18}$.
The data is also used to fit the $\Lambda$CDM model, the results are $\chi^2=160.51$,
$\Omega_m=0.30\pm 0.03$ and $\Omega_k=-0.02\pm 0.02$.
This model fits the observational data as well as the $\Lambda$CDM and the two-parameter models do.
The contours of $\Omega_m$ and $\Omega_k$
are shown in Figs. \ref{fig8} and \ref{fig9}. The errors
on $\Omega_m$ and $\Omega_k$ are almost the same. Comparing with the results in \cite{gong05},
we find that the new SN Ia data improves the constraint on $\Omega_k$ significantly.

\section{Discussion}

By fitting the simple two-parameter representation of $q(z)$ to the new 182 gold
SN Ia data, we find strong evidence of acceleration in the recent past which is
consistent with previous studies in \cite{sturner,gong06}. While the evidence
of current acceleration is weak from fitting the simple piecewise constant acceleration
model to the previous gold SN Ia data \cite{sturner} and fitting the simple two-parameter
representation of $q(z)$ to the 115 nearby Supernova Legacy Survey (SNLS) SN Ia data \cite{astier,gong06},
we find strong evidence of current acceleration by using the gold SN Ia data. The strongest evidence
of acceleration again happens around the redshift $z\sim 0.2$. The transition
redshift when the Universe underwent the transition from deceleration to acceleration
is found to be $z_t=0.36^{+0.23}_{-0.08}$ at the $1\sigma$ level.

The new SN Ia data, together with the BAO measurement from SDSS and
the shift parameter determined from WMAP3 data, are used to fit
three popular DE parameterizations. When we are given the
parameterizations $w(z)=w_0+w_a z/(1+z)$ and $w(z)=w_0+w_a
z/(1+z)^2$, the explicit analytical expressions for the DE density
can be derived. Alternatively, we can think that we are
parameterizing the DE density $\Omega_{DE}(z)$ instead of $w(z)$.
The new observational data makes slightly improvement on the
constraint of $w_a$, while the $\Lambda$CDM model is still
consistent with current observational data. Although high redshift
SN Ia data provides robust constraint on the property of DE
\cite{evldh}, the stringent constraint on $w(z)$ happens around
$z\sim 0.3$. In other words, at the $3\sigma$ confidence level,
$w(z)$ is best constrained around the redshift $z\sim 0.3$.
Surprisingly, we only have a few SN Ia with redshift around $0.3$ in
the current 182 gold SN Ia data. The same result holds for the DE
parametrization
$\Omega_{DE}(z)=1-\Omega_m-A_1-A_2+A_1(1+z)+A_2(1+z)^2$, although
the redshift is now around $z\sim 0.4$. We think this result is
quite generic for two-parameter parametrizations. The result also
suggests that more SN Ia data with the redshift $z\sim 0.2-0.4$ may
be valuable to give strong constraint on $w(z)$.
The sweet spot around the redshift $z\sim 0.2-0.4$ may be argued
from the Hubble law and the decreasing importance of DE \cite{huterer,weller,saini}:
(1) At low redshift, the luminosity distance can be expressed as
$H_0 d_L(z)=z+\frac{1}{2}(1-q_0)z^2$. To the linear approximation, it does not
depend on the cosmological parameters, so the constraint on the property
of DE at low redshift from SN Ia data is not strong; (2) At high redshift,
the role of DE diminishes.
Depending on the model, the evidence for $w(z)<0$ at high redshift is different.

We also apply the one-parameter parametrization $w(z)=w_0\exp(z/(1+z))/(1+z)$ to study
the geometry of the Universe. Although the SN Ia data alone does not provide valuable
constraint on the geometry, the new combined data improves the constraint on $\Omega_k$
significantly. At the $3\sigma$ confidence level, we have $|\Omega_k|\alt 0.05$
for the model $w(z)=w_0\exp(z/(1+z))/(1+z)$
and $-0.064<\Omega_k<0.028$ for the $\Lambda$CDM model.
It should be stressed that the effect of the heterogeneous nature of the gold SN Ia
data on the systematics is also important and it may impose potential problem when
combining with WMAP3 data \cite{jbp06,nesseris05,nesseris6b}. The more homogeneous
SNLS SN Ia data avoids this problem \cite{jbp06,nesseris05,nesseris6b}.

\begin{acknowledgments}
Y.G. Gong is supported by Baylor University, NNSFC under grant No. 10447008 and 10605042,
SRF for ROCS, State Education Ministry
and CMEC under grant No. KJ060502.
\end{acknowledgments}

%\section*{References}

\end{document}